\documentclass[aps,reprint,10pt,notitlepage,pra,
               twocolumn,
               superscriptaddress,
               amsfonts,amssymb,amsmath,
               showpacs,nobibnotes,
               nofootinbib,
               ]{revtex4-1}

\usepackage{graphicx}
\usepackage{amssymb}
\usepackage{amstext}
\usepackage{amsmath}
\usepackage{amsthm}
\usepackage[warn]{mathtext}
\usepackage{braket}
\usepackage{bm}

\usepackage{indentfirst}
\usepackage{wrapfig}
\usepackage{topcapt}
\usepackage{comment}
\usepackage{subfigure}
\usepackage{color,xcolor}
\definecolor{dark-blue}{rgb}{0,0,0.875}
\definecolor{dark-green}{rgb}{0,0.625,0}
\definecolor{dark-red}{rgb}{0.875,0,0}
\usepackage[colorlinks=true,linkcolor=dark-red,filecolor=dark-blue,
citecolor=dark-green,urlcolor=dark-blue,breaklinks=true]{hyperref}
\begin{document}
\title{Influence of classic noise on entangled state formation in parametric systems}
\author{V. O. Martynov}
\affiliation{Institute of Applied Physics of RAS, Nizhny Novgorod, Russia}
\author{V. A. Mironov}
\affiliation{Institute of Applied Physics of RAS, Nizhny Novgorod, Russia}
\author{L. A. Smirnov}
\affiliation{Institute of Applied Physics of RAS, Nizhny Novgorod, Russia}
\affiliation{Department of Control Theory, Nizhny Novgorod State University,
Gagarin Av. 23, 606950, Nizhny Novgorod, Russia}
\begin{abstract}
A study of ``high temperature'' entangled states in a system of two parametrically coupled quantum oscillators placed into independent thermal baths is performed taking into account partially coherent parametric pump. Processes in an open system are considered based on the Heisenberg-Langevin formalism. We obtain a closed system of equations for the averaged quadratic correlation functions in quantum stochastic problem as a result of Markov processes approximation. On the basis of numerical calculations the dynamics of the logarithmic negativity, which is the measure of entanglement in the system, is investigated. It is shown that the partial coherence of the parametric pump makes the lifetime of the entangled states finite. The threshold characteristics of the formation and existence of these states are specified.   
\end{abstract}
\maketitle
\section{Introduction}\label{sec:Introduction}
According to the superposition principle underlying the basis of quantum mechanics, a many-particle system may be found in a so-called entangled state~\cite{nielsen_quantum_2000}. At the present time, the entangled states of several quantum-mechanical objects are actively considered as a special nonlocal resource which can be effectively used for solving a broad range of quantum-optical problems dealing with quantum information transfer and quantum calculations~\cite{nielsen_quantum_2000, bouwmeester_physics_2000, weedbrook_gaussian_2012} (including cryptography~\cite{nielsen_quantum_2000, weedbrook_gaussian_2012, horodecki_quantum_2009} and metrology~\cite{giovannetti_advances_2011, joo_quantum_2011, anisimov_quantum_2010}). The main obstacle on the way of practical application of this resource is decoherence~\cite{bouwmeester_physics_2000, scully_quantum_1997}. In the general case, the loss of quantum coherency (and, as a result, destruction of entangled states) is caused by interaction of subsystems with an environment possessing a great number of degrees of freedom (for example, thermal bath). It is worth noting that the rate of a decoherence process increases together with the rise of temperature~\cite{bouwmeester_physics_2000}.\par
In recent years, the possibility of retaining the entanglement during a long time at finite temperatures in the active systems due to the constant energy inflow has been  actively discussed~\cite{vedral_quantum_2010}. In particular, in Refs.~\cite{galve_bringing_2010, roque_role_2013, schmidt_almost_2013, chen_transition_2015} the problems of two coupled oscillators under the conditions of parametric instability development have been considered. The given physical model enables one to examine the process of the entangled states formation in even more complicated situations, for example, microwave superconducting resonators~\cite{wang_deterministic_2011}, nanomechanics oscillators~\cite{woolley_nonlinear_2008}, and optomechanical systems~\cite{vitali_optomechanical_2007}. Besides, it should be noted that at present analogous processes are observed in biological macromolecules under the action of the short laser pulses~\cite{romero_quantum_2014,halpin_two-dimensional_2014,fuller_vibronic_2014}.\par
In Ref~\cite{galve_bringing_2010} the case when the parametric pumping is realized by variation of the coefficient of linear interaction between oscillators, each placed into its own bath, has been researched. In the paper the existence of the entanglement within the temperature range from zero to the critical value, which may be rather high and determined by the instability increment, has been demonstrated. In Ref~\cite{roque_role_2013} an analogous system with oscillators placed into one common bath has been studied, and the occurrence of entanglement at relatively high temperatures has been shown too. In Ref~\cite{schmidt_almost_2013} the occurrence of the entangled state of two oscillators under the parametric action upon only one of them (in the absence of linear interaction between them, but with mediate connection through the non-Markovian bath) has been discussed.\par
In contrast to the papers~\cite{galve_bringing_2010, roque_role_2013, schmidt_almost_2013} enumerated above, in our work peculiarities of high-temperature entangled states formation under the conditions of partially coherent parametric pumping are considered. The presence of an additive noise connected with pumping has an essential influence on statistical processes in the classical parametric systems~\cite{mandel_optical_1995}. In the quantum case such noise leads to a noticeable efficiency decrease of non-classical (for example, squeezed) atomic states generation~\cite{scully_quantum_1997, yu_sudden_2006}. The study of entanglement for two oscillators coupled parametrically in the situation similar to that discussed in~\cite{galve_bringing_2010}, but taking into account the partial coherency of pumping, is performed on the basis of Heisenberg-Langevin formalism~\cite{scully_quantum_1997}. In this case, according to the rotating wave approximation for Markovian processes, one succeeds in obtaining a closed set of equations for the averaged  quadratic combinations of both the creation and the annihilation operators and in calculating the logarithmic negativity to characterize the entanglement degree of oscillators states.\par
\section{Description of the open dissipative system under partially coherent pumping}\label{sec:Theory}
\subsection{The model and the basic approximations}\label{sec:Model}
Let's consider two linearly coupled identical quantum harmonic oscillators with the same eigen frequencies $\omega_{1}=\omega_{2}=\omega$ and with corresponding annihilation (creation) operators $\hat{a}^{\phantom{\dagger}}_{1}${\,}($\hat{a}^{\dagger}_{1}$) and $\hat{a}^{\phantom{\dagger}}_{2}${\,}($\hat{a}^{\dagger}_{2}$). By analogy with the problem investigated in the paper~\cite{galve_bringing_2010} we shall suppose that each oscillator interacts with its own environment having a great number of freedom degrees, i.{\,}e., it is placed into an isolated thermal bath. In this case evolution of the whole system is determined by the Hamiltonian $\hat{H}$, which may be written in the form of three terms:
\begin{equation}\label{eq:Htotal}
	\hat{H}=\hat{H}_{S}+\hat{H}_{B}+\hat{H}_{I}.
\end{equation}
Here 
\begin{equation}\label{eq:Hsystem}
	\hat{H}_{S}\!=\!\sum_{j=1}^{2}\hbar\omega\!\left(\hat{a}^{\dagger}_{j}\hat{a}^{\phantom{\dagger}}_{j}\!+\!\frac{1}{2}\right)+\frac{\hbar\omega}{2}c\left(t\right)\!\left(\hat{a}^{\phantom{\dagger}}_{1}\!+\!\hat{a}^{\dagger}_{1}\right)\!\!\left(\hat{a}^{\phantom{\dagger}}_{2}\!+\!\hat{a}^{\dagger}_{2}\right)
\end{equation}
presents the Hamilton operator for the subsystem of two oscillators with the coupling coefficient $c\left(t\right)$ depending on time $t$. Further, let's choose the function $c\left(t\right)$ varying nearly harmonically
\begin{equation}\label{eq:coupling}
	c\left(t\right)=\varepsilon\cos\bigl(\varOmega t-\varphi\left(t\right)\bigr).
\end{equation}
with the amplitude of oscillations $\varepsilon$ and the frequency $\varOmega$, close to the value of $2\omega$ satisfying the condition of parametric instability development in the classical problems~\cite{landau_mechanics_2007}, namely, $\left|\varOmega\!-\!2\omega\right|\!\ll\!\omega$. However, unlike~\cite{galve_bringing_2010} where the ideal situation of coherent monochromatic action was considered, we shall take into account the phase fluctuations $\varphi\left(t\right)$ in the coupling coefficient $c\left(t\right)$, which occurs naturally in the real situations. Besides, we shall suppose that the random phase $\varphi\left(t\right)$ has a zero mean value and presents a Wiener process, whose derivative (in general sense) $\dot{\varphi}\left(t\right)$ is a normal white noise with the spectral width $D$, i.{\,}e., the following relations~\cite{scully_quantum_1997}
\begin{equation}\label{eq:ClassicCor}
	\langle\varphi\rangle_{\varphi}=0,
	\hspace{5mm}\langle\dot{\varphi}\left(t\right)\dot{\varphi}\left(t'\right)\rangle_{\varphi}=2D\delta\left(t-t'\right),
\end{equation} 
are satisfied. It should be noted that in~(\ref{eq:ClassicCor}) averaging is performed over the ensemble of realizations $\varphi$. Hereinafter, a dot over the function or operator will denote the time derivative $t$.\par
We shall suppose that the thermal baths represent two ($j=1,2$) independent ensembles consisting of a great number of simple harmonic oscillators. These oscillators, making no contact with each other, have frequencies $\omega_{jk}$, respectively, and are characterized by the annihilation (creation) operators $\hat{b}^{\phantom{\dagger}}_{jk}$ ($\hat{b}^{\phantom{\dagger}}_{jk}$), satisfying the boson commutation relations. Thus, the second term $\hat{H}_{B}$, standing on the right-hand side  of the relation~(\ref{eq:Htotal}) being in essence the Hamilton operator of the aggregate environment for the dedicated (by us) subsystem, will be written as follows
\begin{equation}\label{eq:Hbath}
	\hat{H}_{B}=\sum_{j=1}^{2}\sum_{k=1}^{+\infty}\hbar\omega_{jk}\left(\hat{b}^{\dagger}_{jk}\hat{b}^{\phantom{\dagger}}_{jk}+\frac{1}{2}\right).
\end{equation}
The last term $\hat{H}_{I}$ in Eq.~(\ref{eq:Htotal}) is nothing more but a Hamiltonian of interaction for the coupled quantum harmonic oscillators with their own thermostats. Exactly $\hat{H}_{I}$ causes the dissipative processes leading to decoherence. In the situation described above $\hat{H}_{I}$ has the following form:
\begin{equation}\label{eq:Hinteraction}
	\hat{H}_{I}=\sum_{j=1}^{2}\sum_{k=1}^{+\infty}\hbar\textsl{g}_{jk}\left(\hat{a}^{\phantom{\dagger}}_{j}+\hat{a}^{\dagger}_{j}\right)\!\left(\hat{b}^{\phantom{\dagger}}_{jk}+\hat{b}^{\dagger}_{jk}\right),
\end{equation}
where the coefficients $\textsl{g}_{jk}$ determine the action degree of the $k$-th mode of the $j$-th bath upon the $j$-th oscillator.\par
The described model may be essentially simplified if one applies the rotating wave approximation~\cite{scully_quantum_1997} widely used in quantum optics. This approximation takes into consideration only resonant terms of the total Hamilton operator $\hat{H}$. In addition, we make an assumption about the Markov nature of the random processes~\cite{scully_quantum_1997}, associated with the presence of thermal baths. Firstly, under weak coupling with thermostats when $\textsl{g}_{jk}\bigl/\omega\bigr.\ll 1$, the considered subsystem interacts actively with its environment only in a narrow spectral band from $\omega-\delta\omega$ to $\omega+\delta\omega$ ($\delta\omega\bigl/\omega\bigr.\ll 1$), i.{\,}e., with simple oscillators having values of $\omega_{jk}$ close to the value of $\omega$. In this case and at small amplitude $\varepsilon$ ($\varepsilon\ll 1$) of pumping $c\left(t\right)$ in the form~(\ref{eq:coupling}) and at the period $\tau_{\varOmega}=2\pi\bigl/\varOmega\bigr.$ of its variation comparable with the half period $\tau_{\omega}=2\pi\bigl/\omega\bigr.$ of eigen oscillations of the small system one may neglect the non-resonance terms of the Hamiltonians $\hat{H}_{I}$ and $\hat{H}_{S}$, proportional respectively to $\hat{a}^{\phantom{\dagger}}_{j}\hat{b}^{\phantom{\dagger}}_{jk}$, $\hat{a}^{\dagger}_{j}\hat{b}^{\dagger}_{jk}$, $\hat{a}^{\phantom{\dagger}}_{1}\hat{a}^{\phantom{\dagger}}_{2}e^{-i\varOmega t}$, $\hat{a}^{\phantom{\dagger}}_{1}\hat{a}^{\dagger}_{2}$, $\hat{a}^{\dagger}_{1}\hat{a}^{\phantom{\dagger}}_{2}$ and $\hat{a}^{\dagger}_{1}\hat{a}^{\dagger}_{2}e^{i\varOmega t}$. Note that in this case it is admitted that all characteristic times of relaxation processes are large in comparison with $\tau_{\omega}$. Secondly, we'll consider that the coefficients $\textsl{g}_{jk}$ for each $k$-th mode of $j$-th bath  from the spectral band with the width $2\delta\omega$ near the point $\omega$ are practically the same and approximately equal to one and the same constant value depending on $\omega$: $\textsl{g}\left(\omega\right)$ ($\textsl{g}\left(\omega\right)\!\bigl/\omega\bigr.\ll1$), i.{\,}e., $\textsl{g}_{jk}\approx\textsl{g}\left(\omega\right)$, if $\omega-\delta\omega\leq\omega_{jk}\geq\omega+\delta\omega$.\par
With a knowledge of the total Hamiltonian $\hat{H}$ of the considered model and using the assumptions described above, it is not difficult to obtain a system containing two coupled Heisenberg-Langevin equations
\begin{equation}\label{eq:HeisenbergLangevinEq}
	\dot{\hat{\mathcal{A}}}_{j}\!=\!i\dfrac{\varDelta}{2}\hat{\mathcal{A}}_{j}\!-\!\dfrac{\varGamma}{2}\hat{\mathcal{A}}_{j}\!-\!\dfrac{i\omega\varepsilon}{4}\hat{\mathcal{A}}_{3-j}^{\dagger}e^{i\varphi}\!+\!\hat{\mathcal{F}}_{j}\!\left(t\right)e^{i\varDelta t}\hspace{2.0mm}\left(j\!=\!1,2\right)\!
\end{equation}
for the slowly varying annihilation operators
\begin{equation}\label{eq:SlowAnnihilationOperator}
	\hat{\mathcal{A}}_{j}\!\left(t\right)=\hat{a}^{\phantom{\dagger}}_{j}\!\left(t\right)e^{i\Omega t/2}\hspace{2.0mm}\left(j\!=\!1,2\right),
\end{equation} 
which may be obtained by eliminating the high-frequency dependence on time from each $\hat{a}^{\phantom{\dagger}}_{j}\!\left(t\right)$. Here $\varDelta=\varOmega-2\omega$ is the difference from the condition of the exact parametric resonance, $\varGamma=2\pi\textsl{g}^{2}\!\left(\omega\right)\mathcal{Q}\left(\omega\right)$ is the damping coefficient characterizing the rate of dissipative processes caused by interaction of each quantum harmonic oscillator with its thermal environment, $\mathcal{Q}\left(\omega\right)$ is the states density in the identical thermostats corresponding to $\omega_{jk}=\omega$. Each $j$-th equation~(\ref{eq:HeisenbergLangevinEq}) contains a noise operator $\hat{\mathcal{F}}_{j}\left(t\right)$, depending on variables of $j$-th bath. Both $\hat{\mathcal{F}}_{j}\left(t\right)$ have a zero mean value, i.{\,}e., $\langle\hat{\mathcal{F}}_{j}\left(t\right)\rangle_{R}\!=\!\langle\hat{\mathcal{F}}_{j}^{\dagger}\left(t\right)\rangle_{R}\!=\!0$, and satisfy the following correlation relationships:
\begin{gather}\label{eq:CorrelationRatio}
	\langle\hat{\mathcal{F}}_{j}\left(t\right)\hat{\mathcal{F}}_{j'}\left(t'\right)\rangle_{R}= 
	\langle\hat{\mathcal{F}}_{j}^{\dagger}\left(t\right)\hat{\mathcal{F}}_{j'}^{\dagger}\left(t'\right)\rangle_{R}=0,{}\\{} 
	\langle\hat{\mathcal{F}}_{j}\left(t\right)\hat{\mathcal{F}}^{\dagger}_{j'}\left(t'\right)\rangle_{R}=\varGamma\left(n_{T}+1 \right) \delta\left(t-t'\right)\delta_{jj'},{}\\{}
	\langle\hat{\mathcal{F}}_{j}^{\dagger}\left(t\right)\hat{\mathcal{F}}_{j'}\left(t'\right)\rangle_{R}=\varGamma n_{T}\,\delta\left(t-t'\right)\delta_{jj'},
\end{gather}
where quantum-mechanical averaging is meant by $\langle\ldots\rangle_{R}$, the indexes $j$ and $j'$ possess values $1$ or $2$, and
\begin{equation}\label{eq:nT}
	n_{T}=\left(e^{\hbar\omega\left/k_{B}T\right.}-1\right)^{-1}
\end{equation}
\looseness=-1 is the mean number of equilibrium bosons (thermal quanta) with the energy of $ \hbar\omega$ at the temperature $T$ and $k_{B}$ is the Boltsman constant. Let us underline the fact that the noise operators $\hat{\mathcal{F}}_{j}\!\left(t\right)$, having correlating properties~(\ref{eq:CorrelationRatio}), are required to preserve the commutation relations $\left[\hat{\mathcal{A}}_{j}\!\left(t\right),\hat{\mathcal{A}}_{j}^{\dagger}\!\left(t\right)\right]\!=\!1$ at any time moment. In fact the presence of both relaxation and noise terms at the same time in Eqs.~(\ref{eq:HeisenbergLangevinEq}) is the manifestation of fluctuation dissipative theorem of the statistic mechanics, according to which the dissipation is always accompanied by fluctuations.\par
According to the relations~(\ref{eq:Htotal}),~(\ref{eq:Hsystem}),~(\ref{eq:Hbath}) and~(\ref{eq:Hinteraction}) the Hamiltonian $\hat{H}$ of the complete system is bilinear with respect to operators $\hat{a}^{\phantom{\dagger}}_{j}$ and $\hat{a}^{\dagger}_{j}$ ($j=1,2$). This implies that if the considered quantum harmonic oscillators initially are in the Gaussian states they will preserve this states in the sequel~\cite{weedbrook_gaussian_2012}. Quantum correlations of Gaussian states depend only on the second-order momenta, i.{\,}e., they are completely characterized by the covariance matrix~$\bm{\sigma}$~\cite{buono_quantum_2010}. For the case when an initially dedicated subsystem is divided into two parts (as in our case), the elements of the matrix $\bm{\sigma}$ are calculated by the formula
\begin{eqnarray}\label{eq:CovarianceMatrix1}
	\sigma_{mn}=\langle\hat{\xi}_{m}\hat{\xi}_{n}+\hat{\xi}_{m}\hat{\xi}_{n}\rangle\bigl/2\bigr.-\langle\hat{\xi}_{m}\rangle\langle\hat{\xi}_{n}\rangle,
\end{eqnarray} 
where indexes $m$ and $n$ take on the values from $1$ to $4$, and $\hat{\xi}_{m}$ and $\hat{\xi}_{n}$ are the corresponding components of the four-dimensional vector $\hat{\bm{\xi}}=\left(\,\hat{q}_{1},\hat{p}_{1},\,\hat{q}_{2},\,\hat{p}_{2}\,\right)^{T}$, whose components are determined in terms of  dimensionless operators of coordinates $\hat{q}_{1}$, $\hat{q}_{2}$ and pulses $\hat{p}_{1}$, $\hat{p}_{2}$ of two interacting elements of the considered small system. For the $j$-th ($j=1,2$) quantum harmonic oscillator, $\hat{q}_{j}$ and $\hat{p}_{j}$ are coupled with the annihilation operator $\hat{a}^{\phantom{\dagger}}_{j}$ and the creation operator $\hat{a}^{\dagger}_{j}$ in the following way:
\begin{equation}\label{eq:qp}
	\hat{q}_{j}=\frac{1}{\sqrt{2}}\left(\hat{a}^{\phantom{\dagger}}_{j}+\hat{a}^{\dagger}_{j}\right),\hspace{2.0mm}\hat{p}_{j}=\frac{i}{\sqrt{2}}\left(\hat{a}^{\dagger}_{j}-\hat{a}^{\phantom{\dagger}}_{j}\right).
\end{equation}
The expression~(\ref{eq:CovarianceMatrix1}) yields the conclusion that $\bm{\sigma}$ may be written in the block representation
\begin{equation}\label{eq:CovarianceMatrix2}
	\bm{\sigma}=
	\begin{pmatrix}
		\bm{\alpha} && \bm{\gamma} \\
		\bm{\gamma}^{T} && \bm{\beta}
	\end{pmatrix},
\end{equation}
where each block $\bm{\alpha}$, $\bm{\beta}$, $\bm{\gamma}$ and $\bm{\gamma}^{T}$ has the same dimension $2\times2$. Here $\bm{\alpha}$ and $\bm{\beta}$ are,correspondingly, covariant matrix of the first and the second oscillators (separately), and their cross-correlations are characterized by the $\bm{\gamma}$ matrix. It should be noted that for the discussed case in the relation~(\ref{eq:CovarianceMatrix1}) the sign $\langle\ldots\rangle$ means both the quantum-mechanical averaging and the averaging over the ensemble of realizations of the random Wiener process $\varphi\left(t\right)$, describing phase fluctuations of partially coherent parametric pumping.\par
\subsection{Logarithmic negativity as a measure of entanglement}\label{sec:LogarithmicNegativity}
Within the framework of the model used here, first of all, we are interested in the features of the formation and keeping up of the high-temperature entangled states of harmonic oscillators coupled parametrically in the presence of the phase fluctuations in the pumping. Let's choose the so-called logarithmic negativity $E_{N}$~\cite{vidal_computable_2002} as a measure characterizing  the entanglement degree of the considered quantum- mechanical objects. In the general case, it is determined for a composite quantum system which is the combination of two arbitrary abstract particles having their individual density matrix $\hat{\rho}_{n}^{(1)}$ and $\hat{\rho}_{n}^{(2)}$ correspondingly. In such a system, the value of $E_{N}$ strictly vanishes at the hybrid separable states~\cite{amico_entanglement_2008}, for which the density operator $\hat{\rho}$ may be presented (not uniquely) in the form of a sum of tensor products $\hat{\rho}_{n}^{(1)}$ and $\hat{\rho}_{n}^{(2)}$:
\begin{equation}\label{eq:SeparableState}
	\hat{\rho}=\sum\limits_{n}\varkappa_{n}\hat{\rho}_{n}^{(1)}\otimes\hat{\rho}_{n}^{(2)},\hspace{2.0mm}\varkappa_{n}>0,\hspace{2.0mm}\sum\limits_{n}\varkappa_{n}=1.
\end{equation}
On the contrary, the nonzero value of $E_{N}$ clearly indicates that the entanglement between two constituent elements occurs, i.{\,}e., we speak about quantum correlations which have no classical analogue. In this connection, in the situation when the number of levels is not bounded above, the value of $E_{N}$ may be arbitrarily large~\cite{buono_quantum_2010}.\par
\looseness=-1 The logarithmic negativity $E_{N}$ is widely used and is frequently mentioned in the literature~(see, for example, \cite{weedbrook_gaussian_2012,schmidt_almost_2013,roque_role_2013,horodecki_quantum_2009,galve_bringing_2010,chen_transition_2015,amico_entanglement_2008}). In the case of the Gaussian states, $E_{N}$ is completely determined by symplectic spectrum $\tilde{\nu}_{m}$ ($m=1,\dots,4$) of the covariance matrix $\tilde{\bm{\sigma}}=\bm{\varLambda}\bm{\sigma}\bm{\varLambda}$, where $\bm{\varLambda}=\mathrm{diag}\bigl[\,1,\,1,\,1,\,-1\,\bigr]$ is the $4\times4$ diagonal operator. Actually,
\begin{equation}\label{eq:LogarithmicNegativity1}
	E_{N}=-\frac{1}{2}\sum\limits_{m=1}^4\log_{2}\Bigl[\min\bigl(1,2\left|\tilde{\nu}_{m} \right|\bigr)\Bigr].
\end{equation}
It is necessary to emphasize that $\tilde{\nu}_{m}$ ($m=1,\dots,4$) coincides with the normal eigenvalues of the
matrix product $-i\bm{J}\bm{\sigma}$, in which 
\begin{equation}
	\bm{J}=
	\begin{pmatrix}
		 \bm{O} && \bm{I} \\
		-\bm{I} && \bm{O}
	\end{pmatrix},
	\hspace{2.0mm}\bm{O}=\begin{pmatrix} 0 && 0 \\ 0 && 0 \end{pmatrix},
	\hspace{2.0mm}\bm{I}=\begin{pmatrix} 1 && 0 \\ 0 && 1 \end{pmatrix}.
\end{equation}\par
As noted in the preceding section, the covariance matrix  $\bm{\sigma}$ can be presented in the block form~(\ref{eq:CovarianceMatrix2}). In this connection by virtue of restrictions applied by both the symmetry property and the uncertainty principle, the determinants of the blocks $\bm{\alpha}$, $\bm{\beta}$, $\bm{\gamma}$ and $\bm{\gamma}^{T}$ are  symplectic invariants~\cite{serafini_symplectic_2004}. That is why the quantum correlations, which are present in the system and which we are interested in, may be expressed in terms of the four quantities
\begin{equation}\label{eq:SymplecticInvariants}
	A=\det{\bm{\alpha}},\hspace{2.0mm}B=\det{\bm{\beta}},\hspace{2.0mm}C=\det{\bm{\gamma}},\hspace{2.0mm}\varSigma=\det{\bm{\sigma}}.
\end{equation}
In particular, it follows from~(\ref{eq:LogarithmicNegativity1}) that the logarithmic negativity $E_{N}$ is nothing more but
\begin{equation}\label{eq:LogarithmicNegativity2}
	E_{N}=\max\left(0,-\log_{2}\bigl[2\tilde{\nu}_{-}\bigr]\right),
\end{equation}
where 
\begin{equation}\label{eq:nu_minus}
	\tilde{\nu}_{-}\!=\!\sqrt{\frac{1}{2}\!\left(\left(A\!+\!B\!-\!2C\right)\!-\!\sqrt{\!\left(A\!+\!B\!-\!2C\right)^2\!-\!4\varSigma}\,\right)}
\end{equation}
is minimal in modulo value among the symplectic spectrum $\tilde{\nu}_{m}$ ($m=1,\dots,4$) of the $\tilde{\bm{\sigma}}$ matrix. The difference of $E_{N}$ from zero means that $\tilde{\nu}_{-}<1\bigl/2\bigr.$. The given inequality may be easily transformed to the necessary and sufficient condition of entanglement presence for the two-particle (or two-mode) Gaussian states formulated by R. Simon~\cite{simon_peres-horodecki_2000}.\par
\looseness=-1 Thus, in order to give a quantitative characteristic and in the same way to study the entanglement dynamics of the parametrically coupled quantum oscillators at the arbitrary time moment $t$,
it is necessary to calculate  the logarithmic negativity $E_N$ using the relations~(\ref{eq:LogarithmicNegativity2}) and~(\ref{eq:nu_minus}). Wherein, it is required to know the behavior of all elements of the covariance matrix $\bm{\sigma}$ depending on $t$. According to~(\ref{eq:CovarianceMatrix1}) and~(\ref{eq:qp}), $\bm{\sigma}$ consists of quadratic combinations $\langle\hat{a}^{\phantom{\dagger}}_{j}\hat{a}^{\phantom{\dagger}}_{j'}\rangle$ and $\langle\hat{a}^{\dagger}_{j}\hat{a}^{\phantom{\dagger}}_{j'}\rangle$ of annihilation operators $\hat{a}^{\phantom{\dagger}}_{j}$ and birth operators $\hat{a}^{\dagger}_{j}$, where indexes $j$ and $j'$ take on values $1$ or $2$. Quantities $\langle\hat{a}^{\phantom{\dagger}}_{j}\hat{a}^{\phantom{\dagger}}_{j'}\rangle$ and $\langle\hat{a}^{\dagger}_{j}\hat{a}^{\phantom{\dagger}}_{j'}\rangle$ can be found for an arbitrary value of $t$ using Heisenberg-Langevin equations~(\ref{eq:HeisenbergLangevinEq}). However, the presence of phase fluctuations $\varphi\left(t\right)$ of partially coherent parametric pumping~(\ref{eq:coupling}) and the necessity of additional (except for quantum-mechanical) averaging over the ensemble of random Wiener process realizations $\varphi\left(t\right)$ create a number of difficulties and lead to significant modifications of the standard procedure. The next section will be devoted to the derivation of a closed system of ordinary differential equations~(ODEs) permitting us to calculate symplectic invariants~(\ref{eq:SymplecticInvariants}) in the presence of classical noise in the time-varying coefficient of coupling between two oscillators~(\ref{eq:coupling}).\par
\subsection{Calculation of covariance matrix elements}\label{sec:Formula}
It follows immediately from the  Heisenberg-Langevin equation~\eqref{eq:HeisenbergLangevinEq} that a four-dimensional vector
\vspace{-1.25mm}
\begin{equation} \hat{\bm{v}}=\left(\hat{\mathcal{A}}_{1},\hat{\mathcal{A}}_{2}^{\dagger}e^{i\varphi},\hat{\mathcal{A}}_{2},\hat{\mathcal{A}}_{1}^{\dagger} e^{i\varphi}\right)^{\!T}\vspace{-1.25mm}
\end{equation}
satisfies the evolution equation of the following form:
\vspace{-1.25mm}
\begin{equation}\label{eq:eqVhat}
\dot{\hat{\bm{v}}}\!\left(t\right)=\bm{K}\hat{\bm{v}}\!\left(t\right)+\dot{\varphi}\!\left(t\right)\bm{L}\hat{\bm{v}}\!\left(t\right)+\hat{\bm{f}}\!\left(t\right).\vspace{-1.25mm}
\end{equation}
Here the vector
\vspace{-1.25mm}
\begin{equation}
\hat{\bm{f}}=\left(\hat{\mathcal{F}}_{1}e^{i\varDelta t},\hat{\mathcal{F}}_{1}^{\dagger}e^{i\left(\varphi-\varDelta t\right)},\hat{\mathcal{F}}_{2}e^{i\varDelta t},\hat{\mathcal{F}}_{1}^{\dagger}e^{i\left(\varphi-\varDelta t\right)}\right)^{\!T}
\end{equation}
is determined by phase fluctuations $\varphi\!\left(t\right)$ of the parametric pumping and by the noise operators $\hat{\mathcal{F}}_{j}\!\left(t\right)$ and $\hat{\mathcal{F}}_{j}^{\dagger}\!\left(t\right)$ ($j=1,2$), $\bm{K}$ and $\bm{L}=\mathrm{diag}\left[\,0,\,-i,\,0,\,-i\right]$, in their turn, represent the $4\times4$ matrix, whose elements are independent of time. Particularly
\begin{equation}
\hspace{-2.0mm}\bm{K}\!=\!\begin{pmatrix}
\bm{M} && \hspace{-2.0mm}\bm{O} \\
\bm{O} && \hspace{-2.0mm}\bm{M}
\end{pmatrix}\!,\hspace{2.0mm}
\bm{M}\!=\!\begin{pmatrix}
\left(i\varDelta\!-\!\varGamma\right)\!\bigl/2\bigr. && \hspace{-2.0mm}-i\omega\varepsilon\bigl/4\bigr. \\ i\omega\varepsilon\bigl/4\bigr. && \hspace{-2.0mm}-\!\left(i\varDelta\!+\!\varGamma\right)\!\bigl/2\bigr.
\end{pmatrix}\!.\!
\end{equation}
In the case when a random phase $\varphi\left(t\right)$ is given by the Wiener-Levi process, for the stochastic linear inhomogeneous equation~\eqref{eq:eqVhat} one can find a general solution~\cite{gardiner_stochastic_2009}:
\vspace{-6.25mm}
\begin{multline}\label{eq:solVhat}
\hat{\bm{v}}\left(t\right)=e^{\left(\bm{K}-D\bm{L}^{2}\right)t+\bm{L}\varphi\left(t\right)}\left[\hat{\bm{v}}\left(0\right)+\phantom{\int\limits_{0}^{t}}\right.{}\\{}
\left.\int\limits_{0}^{t}\!e^{-\left(\bm{K}-D\bm{L}^{2}\right)t'-\bm{L}\varphi\left(t'\right)}\hat{\bm{f}}\left(t'\right)\textsl{d}t'\right]\!,
\end{multline}\vspace{-2.5mm}\\
where $\hat{\bm{v}}\!\left(0\right)$ is constructed of corresponding Schroedinger operators $\hat{\mathcal{A}}_{j}\!\left(0\right)$ and $\hat{\mathcal{A}}_{j}^{\dagger}\!\left(0\right)$ $\left(j\!=\!1,2\right)$. Note that when the thermodynamically equilibrium state is chosen as the initial state, from the expression~\eqref{eq:solVhat} with account of $\langle\hat{\mathcal{F}}_{j}\!\left(t\right)\rangle\!=\!\langle\hat{\mathcal{F}}_{j}^{\dagger}\!\left(t\right)\rangle\!=\!0$ it follows that quantum-stochastic mean values of Heisenberg operators $a_{j}\!\left(t\right)$ and $a_{j}^{\dagger}\!\left(t\right)$ are equal to zero at any moment $t$, i{\,}e.{\,}$\langle\hat{a}_{j}\!\left(t\right)\rangle\!=\!\langle\hat{a}_{j}^{\dagger}\!\left(t\right)\rangle\!=\!0$.\par
In order to calculate the averaged quadratic quantities $\langle\hat{a}^{\phantom{\dagger}}_{j}\hat{a}^{\phantom{\dagger}}_{j'}\rangle$ and $\langle\hat{a}^{\dagger}_{j}\hat{a}^{\phantom{\dagger}}_{j'}\rangle$, we shall also use the equalities~\eqref{eq:HeisenbergLangevinEq}, from which for the components of three-dimensional vectors
\begin{gather}
\hat{\bm{u}}_{1}\!=\!\left(\hat{\mathcal{A}}_{1}\hat{\mathcal{A}}_{2}e^{-i\varphi}\!,
\frac{1}{2}\!\left(\!\hat{\mathcal{A}}_{1}^{\dagger}\hat{\mathcal{A}}_{1}\!+\!\hat{\mathcal{A}}_{2}^{\dagger}\hat{\mathcal{A}}_{2}\!\right)\!,
\hat{\mathcal{A}}_{1}^{\dagger}\hat{\mathcal{A}}_{2}^{\dagger}e^{i\varphi}\right)^{\!T}\!,{}\\{}
\hat{\bm{u}}_{2}\!=\!\left(\hat{\mathcal{A}}_{2}^{2}e^{-i\varphi}\!,
\hat{\mathcal{A}}_{1}^{\dagger}\hat{\mathcal{A}}_{2},
\hat{\mathcal{A}}_{1}^{\dagger 2}e^{i\varphi}\right)^{\!T}\!,{}\\{}
\hat{\bm{u}}_{3}\!=\!\left(\hat{\mathcal{A}}_{1}^{2},
\hat{\mathcal{A}}_{1}^{\dagger}\hat{\mathcal{A}}_{2}e^{i\varphi}\!,
\hat{\mathcal{A}}_{1}^{\dagger 2}e^{2i\varphi}\right)^{\!T}\!,{}\\{}
\hat{\bm{u}}_{4}\!=\!\hat{\bm{u}}_{1}e^{i\varphi},\hspace{2.0mm}\hat{\bm{u}}_{5}\!=\!\hat{\bm{u}}_{2}e^{i\varphi}
\end{gather} 
one may obtain five independent close systems of equations, respectively:
\begin{equation}\label{eq:quadratic}
\dot{\hat{\bm{u}}}_{n}\!\left(t\right)=\bm{V}\hat{\bm{u}}_{n}\!\left(t\right)+\dot{\varphi}\left(t\right)\bm{W}_{n}\hat{\bm{u}}_{n}\!\left(t\right)+\hat{\bm{r}}_{n}\!\left(t\right).
\end{equation}
Here $n\!=\!1\ldots 5${\,}, $\bm{V}$ and $\bm{W}_{n}$ are the $3\times3$ matrix with constant in time elements having the following form:
\begin{gather}
\bm{V}=
\begin{pmatrix}
i\varDelta-\varGamma && -i\omega\varepsilon\bigl/2\bigr. && 0 \\
i\omega\varepsilon\bigl/4\bigr. && -\varGamma && -i\omega\varepsilon\bigl/4\bigr. \\
0 && i\omega\varepsilon\bigl/2\bigr. &&  -i\varDelta-\Gamma
\end{pmatrix},{}\\{}
\bm{W}_{1}=\bm{W}_{2}=\mathrm{diag}\left[\,-i,\,0,\,i\,\right],{}\\{}
\bm{W}_{3}=\bm{W}_{4}=\bm{W}_{5}=\mathrm{diag}\left[\,0,\,i,\,2i\,\right].
\end{gather} 
By analogy with $\hat{\bm{f}}\left(t\right)$ vectors $\hat{\bm{r}}_{n}\left(t\right)$ are determined by the phase fluctuations $\varphi\left(t\right)$ of the parametric pumping~\eqref{eq:coupling} and by the noise operators $\hat{\mathcal{F}}_{j}\!\left(t\right)$ and $\hat{\mathcal{F}}_{j}^{\dagger}\!\left(t\right)$, more exactly, by various combinations arising as a result of multiplication of $\hat{\mathcal{F}}_{j}\!\left(t\right)$ (or $\hat{\mathcal{F}}_{j}^{\dagger}\!\left(t\right)$) and $\hat{\mathcal{A}}_{j'}\!\left(t\right)$ (or $\hat{\mathcal{A}}_{j'}^{\dagger}\!\left(t\right)$), where $j,j'=1,2$. In view of the fact that the expressions for the components of given vectors are too lengthy, we shall not list them in the exact form, but  only write their mean values for each of $\hat{\bm{r}}_{n}\!\left(t\right)$:
\vspace{-2.5mm}
\begin{gather}
\langle\hat{\bm{r}}_{1}\!\left(t\right)\rangle=\left(\,-i\omega\varepsilon\bigl/4\bigl.,\,\varGamma n_{T},\,i\omega\varepsilon\bigl/4\bigl.\,\right)^{T}\!,{}\\{}
\langle\hat{\bm{r}}_{2}\!\left(t\right)\rangle=\langle\hat{\bm{r}}_{3}\left(t\right)\rangle=\langle\hat{\bm{r}}_{5}\left(t\right)\rangle=\left(\,0,\,0,\,0\,\right)^{T}\!,{}\\{}
\langle\hat{\bm{r}}_{4}\!\left(t\right)\rangle=\left(\,i\omega\varepsilon e^{-Dt}\bigl/4\bigr.,\,\varGamma n_{T}e^{-D t},\,-i\omega\varepsilon e^{-Dt}\bigl/4\bigr.\,\right)^{T}\!.
\end{gather}\vspace{-3.75mm}\\
When performing the calculations of these mean values we have employed both the general solution~\eqref{eq:solVhat} of Eq.~\eqref{eq:eqVhat} and correlation relations~\eqref{eq:CorrelationRatio}. Besides, we have assumed that $\hat{\mathcal{F}}_{j}\!\left(t\right)$ and $\hat{\mathcal{A}}_{j}\!\left(0\right)$ are statistically independent.\par
Basing on the formalism of stochastic differential equations and the properties of the Ito stochastic integrals~\cite{gardiner_stochastic_2009}, one can show that for a random variable $\varphi\left(t\right)$, governed by Wiener process and satisfying conditions~\eqref{eq:ClassicCor}, the sequential averaging of the expressions~\eqref{eq:quadratic} over ensemble of phase fluctuations $\varphi\left(t\right)$ of parametric pumping and over quantum-mechanical noise, caused by the interaction of a small system of two coupled oscillators with the environment, leads to the following exact equations
\begin{equation}\label{eq:eqUaverage}
\dot{\langle\bm{u}_{n}\rangle}=\left(\bm{V}+D\bm{W}_{n}^{2}\right)\langle\hat{\bm{u}}_{n}\rangle+\langle\hat{\bm{r}}_{n}\rangle\hspace{2.0mm}\left(n=1\ldots 5\right).
\end{equation}
In total, considering each of these relations as a set of the third order ODEs with respect to the components of the vector $\langle\hat{\bm{u}}_{n}\!\left(t\right)\rangle$ and solving it, for example, with the help of one of the Runge-Kutta methods~\cite{garcia_numerical_2000}, it is easy to find mean values of the operators  $\hat{\mathcal{A}}^{\phantom{\dagger}}_{j}\hat{\mathcal{A}}^{\phantom{\dagger}}_{j'}$ and $\hat{\mathcal{A}}^{\dagger}_{j}\hat{\mathcal{A}}^{\phantom{\dagger}}_{j'}$ ($j,j'=1,2$), allowing to restore the behavior of covariance matrix elements $\bm{\sigma}$ and to determine the desired dependence $E_{N}\!\left(t\right)$.\par
\vspace{-1.25mm}
\section{Logarithmic negativity dynamics}\label{sec:Results}
\vspace{-2.5mm}
In this section, we present the results of numerical analysis of the behavior of the logarithmic negativity $ E_{N}\!\left( t\right)$ during the evolution of the system on the basis of Eqs.~\eqref{eq:eqUaverage}. The aim of the study is to demonstrate the effects arising from the accounting for the partial coherence and their impact on the generation and maintenance of entanglement in a system of two parametrically coupled oscillators in the presence of interaction with the environment. First of all, note that according to calculations, the behavior of $E_{N}\!\left(t\right)$ does not vary qualitatively till the value of detuning $\varDelta=\varOmega-2\omega$ lies within the interval $\left(-\varDelta_{\ast},\varDelta_{\ast}\right)$, where $\varDelta_{\ast}=\sqrt{\omega^{2}\varepsilon^{2}-4\varGamma^{2}}\bigl/2\bigr.$, i.{\,}e., the conditions under which parametric instability develops~\cite{landau_mechanics_2007}. Further, for the sake of certainty, we shall limit ourselves to discussion of the situation with the exact resonance, when $\varOmega=2\omega$; in this case the process of formation of entangled states for two quantum harmonic oscillators is most effective. It should be also noted that all the figures given below (for the convenience of their comparison) are plotted at the fixed ratio $Q=\omega\bigl/\varGamma\bigr.=5000$, which in fact is a quality factor of the oscillators. Thus, a certain relation between the eigen frequency $\omega$ and the damping coefficient $\varGamma$ is established, and in the relations~\eqref{eq:eqUaverage} spectral width $D$, detuning $\varDelta$ and time $t$ may be normalized to $\omega$. When solving ODEs~\eqref{eq:eqUaverage}, mean square quantities corresponding to the non-entangled thermodynamically equilibrium states are used as the initial conditions.\par
\begin{figure}[t]
	\begin{minipage}[t]{1.0\linewidth}
	\includegraphics[scale=1.0]{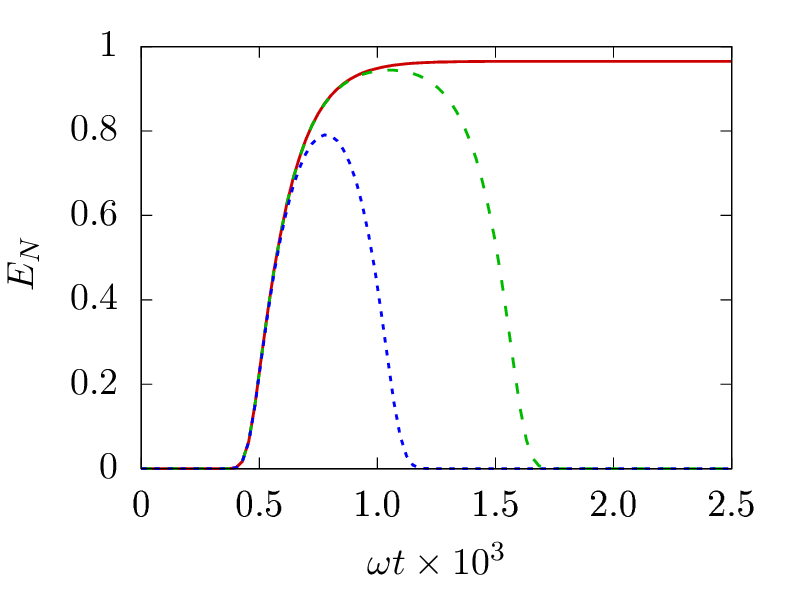}\vspace{-1.25mm}
	\caption{(Color online) The dynamics of logarithmic negativity $E_{N}\left(t\right)$
	characterizing the entanglement degree of two oscillators coupled  parametrically in the presence of interaction with the environment (in the form of two independent thermal baths) is presented for three values of phase-induced spectral width of the pumping line: $D\bigl/\omega\bigr.=0$ (red solid line), $D\bigl/\omega\bigr.=10^{-10}$ (green dashed line), $D\bigl/\omega\bigr.=10^{-8}$ (blue dotted line). For each of these cases $Q=\omega\bigl/\varGamma=5000$, the temperature $T$ of the system corresponds to the mean number of the equilibrium bosons $n_{T}=10$ , and the amplitude of coupling force $c\left(t\right)$ is equal to $\varepsilon=1.6\cdot10^{-2}$.}\vspace{-2.5mm}
	\label{plot:dynamics}
	\end{minipage}
\end{figure}
For the coherent pumping, i.{\,}e., when $D=0$, the results of our calculations agree with the conclusions presented in the paper~\cite{galve_bringing_2010} where they are obtained (including non-Markovian approximation) using the description of small subsystem by dint of the density matrix formalism. Dynamics of the logarithmic negativity $E_{N}\left(t\right)$ in this situation is shown in Fig.~\ref{plot:dynamics} by a red solid line. One can see that entanglement between two coupled oscillators occurs only after some time, depending on the environment temperature $T$ (which means that it also depends on the number of equilibrium bosons $n_{T}$). Later $E_{N}$ increases monotonically in time and finally goes to the non-zero steady-state value $E_{N}^{st}=E_{N}\left(t\to\infty\right)$, which points out the arbitrarily long-term existence of entanglement in the considered system. To form such states the amplitude $\varepsilon$ of coupling coefficient $c\left(t\right)$ needs to be greater than the threshold one $\varepsilon_{0}=4n_{T}\bigl/Q\bigr.$. In other words, there exists a limiting number ${n_{T}}_{0}=Q\varepsilon\bigl/4\bigr.$ of equilibrium bosons in the baths, determining the critical temperature $T_{0}$. At the values of $T$ higher than critical, the entanglement never occurs. At the given $\varepsilon$ the final degree of oscillators entanglement, characterized by $E_{N}^{st}$, increases when $T$ ($T<T_{0}$) decreases. This occurs due to reducing of decoherence effects caused by the influence of thermal noise. In turn, at the constant $T$ the larger amplitude $\varepsilon$ corresponds to the larger value $E_{N}^{st}$.\par
\begin{figure}[t]
	\begin{minipage}[t]{1.0 \linewidth}
	\includegraphics[scale=1.0]{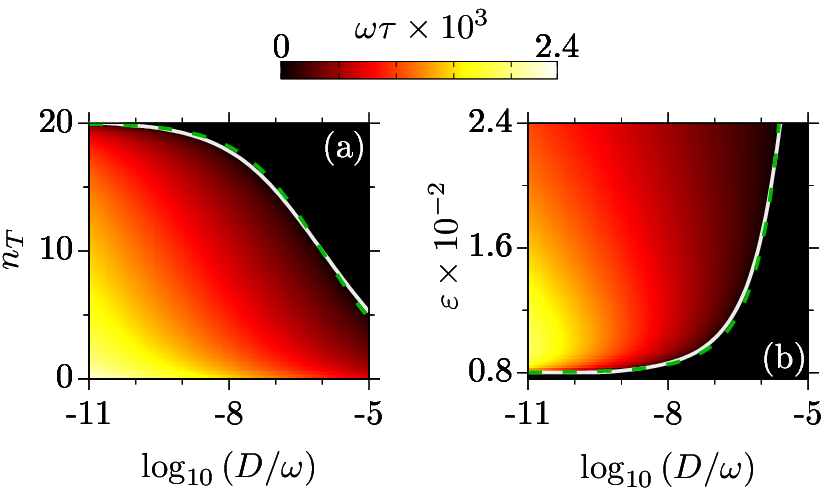}\vspace{-1.25mm}
	\caption{(Color online) (a) The normalized to frequency $\omega$ lifetime $\tau$ of the entanglement states of two quantum harmonic oscillators coupled parametrically depending on the spectral width $D$ of the pumping noise and the number $n_{T}$ of equilibrium bosons in the baths at the given amplitude $\varepsilon=1.6\cdot 10^{-2}$ of the coupling coefficient $c\left(t\right)$. (b) The normalized to frequency $\omega$ lifetime $\tau$ of the entanglement in the considered system depending on $D$ and $\varepsilon$ at the fixed temperature $T$, corresponding to $n_{T}=10$. White solid line shows the interface between the domains with $\tau\neq0$ and $\tau=0$. Green dashed line presents the analytic approximation~\eqref{eq:boundary}.}\vspace{-2.5mm}
	\label{plot:EntTime:DifferentNoise}
	\end{minipage}
\end{figure}
As one can see from the performed within the framework of Eqs.~\eqref{eq:eqUaverage} numerical calculations, taking into account the phase fluctuations $\varphi\left(t\right)$ of partially coherent pumping $c\left(t\right)$ leads to the situation when the arising (as a result of parametric instability development) high-temperature entanglement between two linearly-coupled quantum harmonic oscillators exists only during a finite time interval. Particularly, it is seen in Fig.~\ref{plot:dynamics}, where the green dashed line and the blue dotted line represent the dynamics of logarithmic negativity $E_{N}\left(t\right)$ at finite values of the spectral width $D$ of the noise of the coupling coefficient~$c\left(t\right)$: $D\bigl/\omega\bigr.=10^{-10}$ and $D\bigl/\omega\bigr.=10^{-8}$. These curves, unlike the red solid line, corresponding to the case $D=0$, not simply increase monotonically attaining the positive stationary value $E_{N}^{st}$, but reach their maximum at some moment and then vanish again, which testifies to the disappearance of nonlocal quantum properties in the considered system. Apparently, such a dependence $E_{N}\left(t\right)$ at $D\neq0$ is explained by the gradual increase of influence of the noise $\varphi\left(t\right)$, being the Wiener process, at the averaged quadratic combinations composed of $\hat{a}_{j}\left(t\right)$ and $\hat{a}_{j}^{\dagger}\left(t\right)$ ($j=1,2$). One might say that a random phase $\varphi\left(t\right)$ of the coupling coefficient $c\left(t\right)$ is passed to the annihilation  operators $\hat{a}_{j}\left(t\right)$ and the creation operators $\hat{a}_{j}^{\dagger}\left(t\right)$ and starts to determine their stochastic properties. Eventually, quantum correlations between two harmonic oscillators being in Gaussian states and described by covariant matrix~\eqref{eq:CovarianceMatrix1} are destroyed. It should be noted that the similar (close to the sense) effects have been discussed in Ref~\cite{yu_sudden_2006}, where a sudden death of the entanglement between particles with spin placed into the fluctuating magnetic field has been shown.\par
\looseness=-1 Let us define the time interval during which the logarithmic negativity $E_{N}\left(t\right)$ is strictly nonzero (i.{\,}e., $E_{N}\left(t\right)>0$) as the lifetime $\tau$ of the entangled states in the considered system. The distribution of $\tau$ depending on the spectral width $D$ of the noise in the pumping, normalized to $\omega$, and on the number $n_T$ of equilibrium bosons in the baths determining the temperature $T$ according to~\eqref{eq:nT} at the given amplitude $\varepsilon$ of $c\left(t\right)$ is presented in Fig.~\ref{plot:EntTime:DifferentNoise}$(\mathrm{a})$. One can see that at the fixed $n_T$ the more $D$, the less $\tau$. For the $D$ exceeding some critical value $D_{0}$, the lifetime $\tau$ vanishes. In this case the entanglement between two considered oscillators never occurs. The similar situation is observed if we increase $n_{T}$, i.{\,}e., raise the temperature $T$ at a chosen~$D$. Note that the number ${n_{T}}_{0}\left(D=0\right)=Q\varepsilon\bigl/4\bigr.$ determines, according to~\eqref{eq:nT}, the maximal limiting temperature  ${T_{0}}_{\max}$, with the entanglement states formation region beneath it. For the case presented in Fig.~\ref{plot:EntTime:DifferentNoise}\,$(\mathrm{a})$, parameters are $\varepsilon=1.6\cdot10^{-2}$, ${n_{T}}_{0}\left(D=0\right)=20$. Here the white solid line starting from the point with the components $D=0$ and ${n_{T}}_{0}=20$ divides the parameters plane $\log_{10}\left(D\bigl/\omega\bigr.\right)$ and $n_{T}$ into two domains: in one of which the lifetime $\tau$ is nonzero and in the other $\tau=0$. For the interface of these domains we have written the analytic approximation:
\begin{equation}\label{eq:boundary}
{n_{T}}_{0}\!=\!\dfrac{Q\varepsilon}{4}\!\left[1\!+\!\left(\left(a_1\sqrt{\varepsilon}\!+\!b_1\right)\!Q\!+\!a_2\varepsilon\!+\!b_2\right)\sqrt{\dfrac{D}{\omega}}\right]^{-1}\!\!\!\!\!,
\end{equation}
where numbers $a_{1}=2.08$, $b_{1}=-4\cdot10^{-2}$, $a_{2}=-1.9\cdot10^{3}$, $b_{2}=-4.82$ are selected to best match the results of numerical simulation. Approximation \eqref{eq:boundary} has a high degree of accuracy up to $D/\omega\lesssim10^{-6}$. This fact is substantiated by Fig.~\ref{plot:EntTime:DifferentNoise}, where the green dashed line corresponds
to the formula~\eqref{eq:boundary}. The given approximation permits us to find the critical temperature $T_{0}$ and the threshold value of amplitude $\varepsilon_{0}$ of oscillations of the coupling coefficient $c\left(t\right)$ at the finite spectral width $D$ of phase fluctuations of the pumping.\par
\begin{figure}[t]
	\begin{minipage}[t]{1.0 \linewidth}
	\includegraphics[scale=1.0]{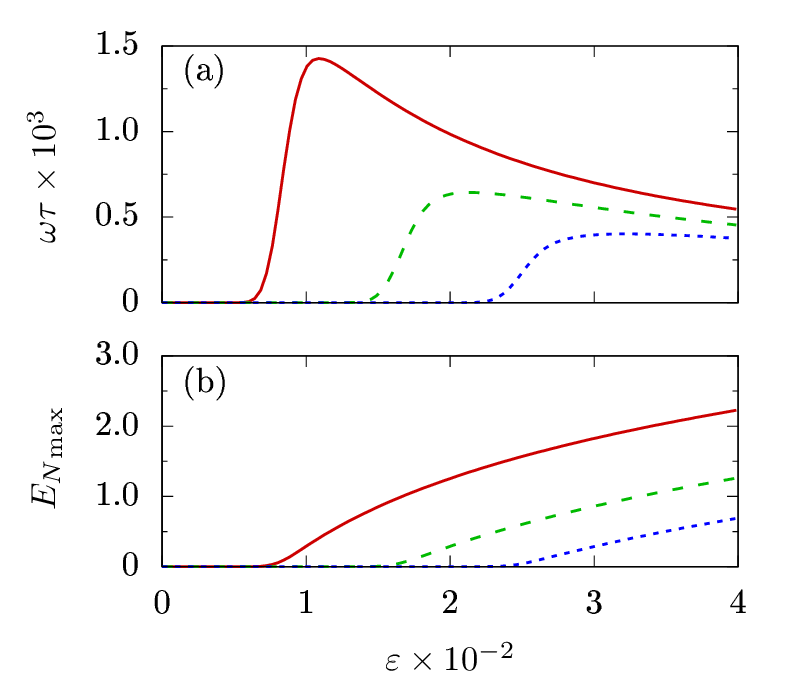}\vspace{-1.25mm}
	\caption{(Color online) $(\mathrm{a})$ The normalized to frequency $\omega$ lifetime $\tau$ of the entanglement between two coupled quantum harmonic oscillators and $(\mathrm{b})$ the maximal value ${E_{N}}_{\max}$ of logarithmic negativity depending on the amplitude $\varepsilon$ of coupling coefficient $c\left(t\right)$ at the fixed spectral width of the noise $D/\omega\!=\!10^{-10}$ in the partially coherent pumping and at three values of the  number of equilibrium bosons in the baths: $n_{T}\!=\!10$ (red solid line), $n_{T}\!=\!20$ (green dashed line) and $n_{T}\!=\!30$ (blue dotted line).}\vspace{-2.5mm}
	\label{plot:EntTime:PumpPower}
	\end{minipage}
\end{figure}
\looseness=-1 The distribution of lifetime $\tau$ of the entanglement of two parametrically coupled quantum harmonic oscillators
depending on $\log_{10}\left(D\bigl/\omega\bigr.\right)$ and $\varepsilon$ at the fixed number $n_T=10$ of equilibrium bosons in baths is shown in Fig.~\ref{plot:EntTime:DifferentNoise}\,$(\mathrm{b})$. Here as well as in the inset $(\mathrm{a})$, the white solid line divides the plane of parameters $\log_{10}\left(D\bigl/\omega\bigr.\right)$ and $\varepsilon$ into two parts, in one of which the lifetime $\tau\neq0$, is nonzero and in the other $\tau=0$. The minimum of the threshold amplitude $\varepsilon_{0}$, exceeding which makes it possible to generate temporarily the entanglement in the considered system, corresponds to $D=0$, and is equal to ${\varepsilon_{0}}_{\min}=4n_{T}\bigl/Q\bigr.=0.8\cdot10^{-2}$. From Fig.~\ref{plot:EntTime:DifferentNoise}{\,}$(\mathrm{b})$, in particular, it follows that for sufficiently coherent pumping the maximal lifetime $\tau$ of the created entangled states is reached close to $\varepsilon_{0}\left(D\right)$. The same can be seen from Fig.~\ref{plot:EntTime:PumpPower}{\,}$(\mathrm{a})$, where the dependences $\tau$ on $\varepsilon$ at the fixed spectral width of the phase fluctuation $D/\omega=10^{-10}$ and at three values of the number of equilibrium bosons in the baths: $n_{T}=10$ (red solid line), $n_{T}=20$ (green dashed line) and $n_{T}=30$ (blue dotted line). However, it is necessary to emphasize that for various practical applications it is important to know not only the duration $\tau$, but also the entanglement degree of the formed states too. It is actually characterized by the maximal value ${E_{N}}_{\max}$, which the logarithmic negativity $E_{N}\left(t\right)$ reaches in the evolution process of the discussed system. In Fig.~\ref{plot:EntTime:PumpPower}{\,}$(\mathrm{b})$ the functions ${E_{N}}_{\max}\left(\varepsilon\right)$ obtained as a result of numerical calculations at the same parameters as the curves from inset $(\mathrm{a})$ are shown. One can see that ${E_{N}}_{\max}$ increases monotonically with the growth of $\varepsilon$. Thus, with the rise of $\varepsilon$, the entanglement becomes more qualitative but less ``long-living''. These facts must be taken into consideration when choosing the optimal amplitude $\varepsilon$ of the harmonic variation of the coupling coefficient $c\left(t\right)$ between the oscillators.\par
\vspace{-0.5mm}
\section{Conclusions}\label{sec:Conclusion}
\vspace{-0.5mm}
Thus, we have studied the peculiarities of the high-temperature entanglement formation under the conditions of partially coherent pumping in a system containing two identical quantum harmonic oscillators coupled linearly, each of them being embedded in a separate independent thermal bath. The discussed problem is important and rather actual from the point of view of creating and maintaining the entangled states in the mesoscopic quantum systems. In conclusion let us summarize our work.\par
In the case of coherent pumping, when the spectral width of the noise equals zero, our results agree with the conclusions made in the paper~\cite{galve_bringing_2010}, where the description of a small system was made by dint of the density matrix formalism. In particular, we have demonstrated using a more intuitive approach that the entangled states with arbitrarily great lifetime and characterized by the stationary logarithmic negativity are formed as a result of parametric instability development (i.{\,}e., monotonic increase of the stored energy) when the coupling coefficient between two oscillators are varied harmonically, even in the presence of thermal baths. In order to form such states it is necessary for the amplitude of pumping oscillations to exceed some critical value proportional to the environment temperature. By analyzing the given relation depending on the pumping amplitude, one can find the limiting temperature, up to which the described above entanglement occurs.\par
For the considered way of pumping the presence of classical phase noise in it (which is natural for the real conditions) leads to the situation when the entanglement between two quantum harmonic oscillators occurs and exists only during a finite time interval. This interval may be easily determined using the condition of the nonzero value of logarithmic negativity, which first increases up to some maximum and then decreases and vanishes according to the performed calculations and within the framework of our approximations. The spectral width of the noise has some threshold value, starting from which the formation of entangled states in the discussed system becomes impossible. The given threshold value increases monotonically with increasing pump amplitude.\par
At the fixed values of the bath's temperature and of the spectral line width of parametric action the maximal lifetime of the entangled states is observed close to the threshold (in pumping amplitude). The degree of entanglement (being defined by the value of the logarithmic negativity) increases monotonically with growing amplitude of the harmonic variation of the coupling parameter between quantum oscillators. This fact should be taken into consideration in various practical applications, when the quality of the entangled states is as important as the time of their existence.\par
Note that in addition to classical phase noise in the pumping, as it was mentioned in papers~\cite{galve_bringing_2010,roque_role_2013}, one of the basic factors destroying the entanglement between two identical oscillators coupled parametrically, may be their nonlinearity, which sooner or later must show itself during the monotonic growth of the stored energy and constrain this process. In our opinion, it will be interesting to determine lifetime of the entangled states, for example, in the problem analogous to that discussed in the given work, but supposing a weak inverse action of composite elements upon independent boson thermostats under the condition of the local in time perturbation, i.{\,}e., in the Markovian approximation. Thereby, we would succeed in comparing the degree of destructive influence of the saturating nonlinearity and partial coherency of pumping on the effect of occurrence and existence of the entangled states in the open systems.\par 
\looseness=-1 
The obtained results may be of interest for the theoretical interpretation of the experiments where the dynamics of biological macromolecules under the action of short laser pulses is studied and the presence of long-living entanglement in the given systems is observed. Note that under natural conditions such biological macromolecules are affected by strong noise. In this situation, the mathematical apparatus developed by us will help to estimate the quality and lifetime of the entangled states.\par
\vspace{-1.0mm}
\begin{acknowledgments}
\vspace{-0.5mm}
This work was supported by the Russian Foundation for Basic Research (project No. 16-32-00750). L.A.S. acknowledges support from the Russian Science Foundation, Grant No. 14-12-00811.\par
\end{acknowledgments}
\bibliographystyle{apsrev4-1}
\bibliography{literature}
\end{document}